\documentstyle[11pt,newpasp,twoside]{article}
\def\edcomment#1{\iffalse\marginpar{\raggedright\sl#1\/}\else\relax\fi}
\reversemarginpar

\begin{document}
\outer\def\gtae {$\buildrel {\lower3pt\hbox{$>$}} \over 
{\lower2pt\hbox{$\sim$}} $}
\newcommand{\Msun} {$M_{\odot}$}

\title{The {\sl XMM-Newton}-MSSL survey of Polars}
 \author{Gavin Ramsay, Mark Cropper}
\affil{Mullard Space Science Lab, University College London, Holmbury
 St.Mary, Dorking, Surrey, RH5 6NT, UK}

\begin{abstract}
The {\sl XMM-Newton}-MSSL survey of polars is a snapshot survey of
nearly 40 polars. We present the initial results, with an emphasis on
the energy balance in these systems and the number of systems which
were observed in a low accretion state. In contrast to the {\sl ROSAT}
results we find that the vast majority of polars show no `soft X-ray
excess'. We find that half of our sample were observed in low or off
accretion states. We find no evidence that these systems were biased
towards certain orbital periods.

\end{abstract}

\section{Introduction}

{\sl XMM-Newton} was launched in Dec 1999. It has the largest
effective area of any X-ray imaging satellite. It was also the first
X-ray mission to have an on-board optical/UV telescope. {\sl
XMM-Newton} has 3 identical X-ray telescopes: one feeds an imaging CCD
(EPIC pn) detector operating from $\sim$0.15--10keV. The other two
telescopes feed both an imaging CCD detector (EPIC MOS,
$\sim$0.15--10keV) and a reflection grating spectrometer (RGS,
$\sim$0.4-2.4keV). The optical/UV telescope (OM) has a diameter of
30cm and has a series of filters extending from the $V$ band down to
$\sim$2000\AA. It has several modes of operation including an imaging
and a fast mode which enables a time resolution less than a second. See
A\&A 2001, Vol 365, No.1, for an overview of the mission and first
science results.

\section{The survey}

We have carried out a survey of polars using {\sl XMM-Newton}.
Currently, 33 polars have been observed as part of this survey, with
another 6 more systems still to be observed. Three systems were
observed as part of the performance verification phase of the
mission. One system, RX J2115--58 was the subject of a series of
pointings (cf Cropper, Ramsay \& Marsh, this volume). Figure 1 shows
the orbital period distribution of all known polars together with
those systems currently observed as part of our survey. The period
distribution of those polars in our survey is comparable with the
observed distribution.

The observation length of each polar was typically
$\sim$5--6ksec. This was long enough so that we obtained full orbital
coverage for many systems. We obtained fast mode observations in the
$V$, UVW1 (2400-3400\AA) and UVW2 (1800-2400\AA) filters using the
optical/UV telescope. Most systems were too faint to obtain useful
spectra using the RGS. There are a great many aspects that such a
survey can address. Here, we discuss the light curves of individual
polars, then provide an initial view of the issue of the `soft X-ray
excess' and examine the implications of the fraction of polars which
were in a low accretion state.

\begin{figure}
\begin{center}
\setlength{\unitlength}{1cm}
\begin{picture}(8,5)
\put(0.,-1.4){\includegraphics{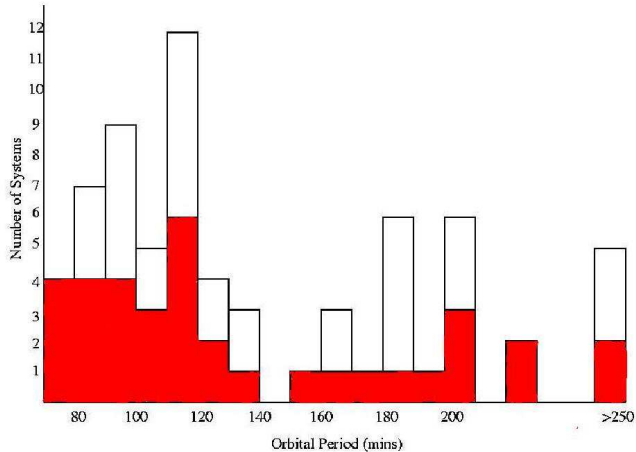}}
\end{picture}
\end{center}
\caption{The orbital period distribution of polars together with those
currently observed as part of the {\sl XMM-Newton}-MSSL polar survey.}
\label{rc_survey} 
\end{figure}

\section{Light curves}

Systems which have only one visible accreting pole are particularly
interesting in that they allow us to probe the structure of the
accretion region as it rotates into view. One example is WW Hor where
we directly measure a change in the X-ray hardness ratio which is due
to the top of the accretion column first coming into view with the
lower layers later (Pandel et al 2002). Simultaneous $B$ band data
using the OM show a quasi-sinusoidal modulation originating from the
photosphere of the white dwarf and near-simultaneous $R$ band data
obtained from SAAO show flux originating from the cyclotron emission.
These data are a particularly good at showing the multi-wave band and
multi-emission processes seen in polars.

Other examples of one-pole systems include EV UMa and GG Leo which
both show prominent accretion stream dips in soft X-rays (and in the
UV in the case of GG Leo). These have allowed us to place constraints
on the radius and total column density in the stream (Ramsay \&
Cropper 2003). All these systems previously mentioned show prominent
flux above 1keV. In complete contrast EU UMa shows a low count rate
above 1keV, and very little evidence for the presence of a shock.

One unusual aspect of the observations of BY Cam was that the hard
X-rays trailed the soft X-rays and are anti-correlated (Ramsay \&
Cropper 2002). This may imply that dense blobs of material impact the
white dwarf and release a sufficient amount of optically thick
material to momentarily obscure the hard X-rays.

Timing studies of the eclipse of DP Leo (Pandel et al 2002) show that
the spin period is slightly shorter than the orbital period and the
orbital period is decreasing much faster than expected for energy loss
by gravitational radiation alone. This is consistent with a separate
study of DP Leo by Schwope et al (2002).

In the next section we discuss the energy balance of AN UMa using {\sl
XMM-Newton} and {\sl ROSAT} data. An example of the general quality of
the {\sl XMM-Newton} X-ray data, we show the soft and hard X-ray light
curves of AN UMa using the EPIC pn detector in Figure 2.

\begin{figure}
\begin{center}
\setlength{\unitlength}{1cm}
\begin{picture}(8,6)
\put(-1,-0.5){\includegraphics{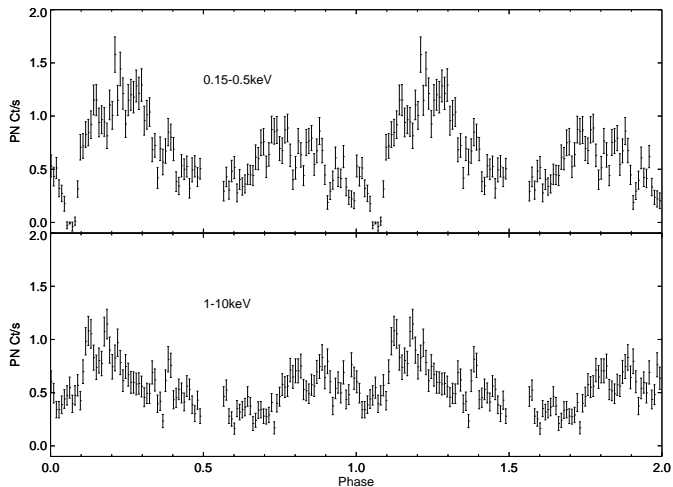}}
\end{picture}
\end{center}
\caption{The X-ray data of AN UMa taken using the EPIC pn of the polar
AN UMa. The top panel shows the soft X-ray light while the lower shows
the hard X-ray light curve.}
\label{rc_anuma} 
\end{figure}

\section{The energy balance}

For polars in a high accretion state the accretion flow forms a strong
shock at some height above the photosphere of the white dwarf. The
maximum temperature in the shock is set by the mass of the white
dwarf. For a 0.7\Msun white dwarf the shock temperature is
$\sim$30keV, with the temperature decreasing as the gas settles onto
the white dwarf. Some fraction of the hard X-rays intercept the
photosphere of the white dwarf, are thermalised and then re-radiated
as soft X-rays or in the extreme UV. The standard model of the shock
region predicts $L_{re-processed}/L_{shock}\sim0.5$ (eg Lamb \&
Masters 1979).

Observations made using {\sl EXOSAT} showed that a number of polars
showed a large `soft X-ray excess': assuming the re-processed
component was emitted as soft X-rays, the ratio,
$L_{re-processed}/L_{shock}$, was well in excess of that predicted by
the standard model. Observations made using {\sl ROSAT} largely
supported this view. Ramsay et al (1994) showed that when the
cyclotron emission was included (which contributes to $L_{shock}$) and
when bolometric unabsorbed luminosities were used, a significant
number of systems showed large soft X-ray excesses. Using unabsorbed
fluxes in the {\sl ROSAT} band (0.1-2.4keV) Beuermann \& Burwitz
(1995) found that all systems showed large excesses. These authors
claimed that for systems with magnetic field strengths \gtae 30MG
cyclotron emission dominates the emission from the shock: if this was
taken into account then the excess would largely disappear. In
contrast, Ramsay et al (1994) did make an estimate of the observed
cyclotron emission and include this in $L_{shock}$ but still found a
significant number of systems with excesses.

To account for this excess, Kuijpers \& Pringle (1982) suggested that
soft X-rays could also be produced by dense blobs of material which
penetrate into the photosphere of the white dwarf without forming a
shock. There is also some evidence that the reprocessed component is
emitted not as soft X-rays but in the extreme UV and therefore any
distinct soft X-ray component is due to blobs of material (Heise \&
Verbunt 1988). The energy balance in polars was therefore one of the
prime areas for study using the {\sl XMM-Newton}-MSSL survey.

\small
\begin{table}
\begin{center}
\begin{tabular}{lr}
\hline
\Large
Source & $L_{soft,bol}/L_{hard,bol}$\\
\hline
EU UMa (RE J1149+28) &  47   \\
V895 Cen             &  13   \\
RX J1007-20          &  10   \\
RE J0501-03          &  3.8  \\
EK UMa               &  1.7  \\
RX J1002-19          &  1.6  \\
EU Cnc               &  1.2  \\
DP Leo               &  1.1  \\
AN UMa               &  1.0  \\
V347 Pav             &  0.7  \\
EV UMa               &  0.5  \\
BY Cam (soft-pole)   &  0.5  \\
GG Leo (RX J1015+09) &  0.5  \\
RX J2115-58 (soft-pole) &  0.3  \\
EP Dra               & 0.2 \\
RX J2115-58 (hard-pole) & -  \\
WW Hor               &  -   \\
CE Gru (Grus V1)     &  -   \\
V349 Pav (V2009--65) &  -   \\
BY Cam (hard-pole)   &  -   \\
V1500 Cyg            &  -   \\
\hline
\end{tabular}
\end{center}
\caption{The ratio $L_{soft,bol}/L_{hard,bol}$ determined for those systems in
a high accretion state.}
\end{table}
\normalsize

In fitting the X-ray spectra we used a multi-temperature emission
model to model the post-shock region (Cropper et al 1999). This is in
contrast to the {\sl ROSAT} studies which used an (inappropriate)
single temperature bremsstrahlung model. Some uncertainty arises as to
the geometrical correction factor which should be applied to
$L_{soft}$ -- this is necessary since this component is optically
thick. We have used a best estimate.

We show in Table 1 the ratio $L_{soft}/L_{hard}$ for our current
sample of polars which were found to be in a high accretion state. We
find that very few systems show high ratios -- only EU UMa, V895 Cen
and RX J1007-20 do. Moreover, even {\sl without} including the
cyclotron component, we find that most systems are roughly consistent
with the standard accretion model. A major surprise was that for 6
systems, at least one pole did not show evidence for a distinct soft
X-ray component -- the X-ray spectrum could be modelled with the
multi-temperature emission model alone.

These results are in contrast to that found in the {\sl ROSAT}
era. There are various reasons for the difference between the {\sl
ROSAT} and {\sl XMM-Newton} results: the {\sl ROSAT} energy band was
not very sensitive to the hard X-ray component; in the {\sl ROSAT} era
a single temperature bremsstrahlung was used to model the hard
component and the resolution of {\sl ROSAT} was lower than {\sl
XMM-Newton} so it was more difficult to constrain the absorption
model. To examine this in more detail, we have reanalysed the {\sl
ROSAT} data of one of the archtypical polars AN UMa (see Fig 2 for its
{\sl XMM-Newton} light curve).

We compare the unabsorbed flux (rather than the luminosity, to make
matters as simple as possible) of the soft and hard components using
different models and satellite. For the multi-temperature shock models
we take the best fit parameters obtained using the {\sl XMM-Newton}
data and fix these (apart from the normalisation) when fitting the
{\sl ROSAT} spectra. The results are shown in Table 2.

We find that if we restrict the energy range to the {\sl ROSAT} band
(as in Beuermann \& Burwitz 1995), the resulting soft/hard ratio is
typically much greater compared to using bolometric fluxes. Further,
for the {\sl ROSAT} data we find that using a one temperature shock
model gives a greater ratio. However, perhaps the most interesting
result is when we compare these ratios with those derived using {\sl
XMM-Newton} spectra. The soft/hard ratio determined over the {\sl
XMM-Newton} energy range and their bolometric fluxes are much reduced
compared to {\sl ROSAT} data using every spectral model. Indeed, when
we take geometrical factors into account to convert to luminosities,
these ratios are consistent with the standard accretion model -- even
when we do not include a contribution from the cyclotron component.
The effect of including a partial neutral absorption model is also
noticeable. In the case of the {\sl XMM-Newton} data the resulting
mass of the white dwarf is lower when we include a partial covering
component. The equivalent absorption column is also greater. Both
these factors contribute to a higher soft/hard ratio.

\setcounter{table}{1}
\begin{table}
\begin{center}
\begin{tabular}{lllll}
\hline
Model & $flux_{bolo,soft}/$ & $flux_{bolo,soft}/$  & 
$flux_{0.1-2.4,soft}/$ &
$flux_{0.15-10,soft}/$\\
      & $flux_{bolo,hard}$ & $flux_{bolo,hard}$ 
      & $flux_{0.1-2.4,hard}$ & 
$flux_{0.15-10,hard}$ \\
    & {\sl ROSAT} & {\sl XMM-Newton} & {\sl ROSAT} & {\sl XMM-Newton}\\
\hline
$wa bb brem$ (30keV) & 9.9 & 1.2 & 22.7 & 0.8\\
$wa bb brem$ (45keV)   & 7.5 & 1.0 & 21.6 & 0.8\\
$wa bb sac$          & 2.3 & 1.4 & 7.7 & 0.9\\
$wa pcf bb sac$      & 5.2 & 3.0 & 7.6 & 1.4\\
\hline
\end{tabular}
\end{center}
\label{rc_rosat}
\caption{The ratio of the unabsorbed flux in the soft and
hard X-ray components derived for AN UMa in {\sl ROSAT} and {\sl
XMM-Newton} using different spectral models. We show the bolometric
flux and the flux over the {\sl ROSAT} and {\sl
XMM-Newton} energy bands. $wa$ refers to a neutral
absorber, $pcf$ a neutral absorber with partial covering, $bb$ a
blackbody, $brem$ a thermal bremsstrahlung, and $sac$ a
multi-temperature accretion column.}
\end{table}

Until now, one of the defining characteristics of polars has been the
presence of a strong soft X-ray component. It is therefore surprising
that we find 6 systems which showed at least one accretion pole which
did not show a distinct soft X-ray component at all. It is likely that
the re-processed component is cool enough to have moved out of the
{\sl XMM-Newton} band. In the scenario of Heise \& Verbunt (1988),
those systems which show a soft component are accreting (at least
some) dense blobs of material. However, for most of these systems the
energy balance is consistent with the standard accretion model --
which does not include blobs. Therefore for most systems where we
detect a distinct soft X-ray component, but the ratio is compatible
with that predicted by the standard model, the reprocessed component
is in soft X-rays rather than the EUV as in the Heise \& Verbunt
scenario. Clearly, we need a better understanding of the spectral
characteristics of the re-processed component. The factors that may
affect its temperature are $M_{wd}$ (since that sets the maximum
temperature in the shock), $\dot{M}$ (since this sets the height of
the shock) and ${\bf B}$ (since this sets amount of cooling due to
cyclotron radiation).

\section{Systems in a low accretion state}

One of the most surprising findings from our survey was that 15 out of
the 33 currently observed polars in our sample were in low or much
reduced accretion rates. Whether a system was in such a state was
determined by comparing the X-ray count rate with previous X-ray
missions or from its long term optical light curve (since we have
simultaneous UV-optical coverage).

To investigate this further we re-examined the fraction of all known
polars that were in a high accretion state at the time of the {\sl
ROSAT} all-sky survey.  Out of the 67 polars known at the end of 2002,
25 were not discovered using the {\sl ROSAT} all-sky survey. Of these
systems 16 were found to be in a low accretion state. This is an even
higher fraction than determined from our {\sl XMM-Newton} survey. The
fact that we observe around half or more polars in a low state is
important when computing population models of polars.

Using our {\sl XMM-Newton} sample, we tested whether those systems in
a low accretion state were biased towards certain orbital
periods. Using a rank test we find no evidence for high or low states
correlating with period or preference. We also tested whether those
systems which were in an off state (as opposed to those systems which
were in low states but detectable) were biased: we find that there is
marginal evidence for a bias towards shorter periods at the 91\%
confidence level.

Of the 15 systems detected in low or off states, 10 were detected in
both the UVW1 and UVW2 filters in the optical/UV telescope. The UV
fluxes remain approximately constant over the orbital cycle. It is
therefore likely that we are detecting the white dwarf photosphere
with little contribution from a heated accretion region. Work is in
progress to determine their temperatures and compare them to
temperatures derived in the high accretion state.

\begin{figure}
\begin{center}
\setlength{\unitlength}{1cm}
\begin{picture}(8,4.5)
\put(-4.,-0.5){\includegraphics{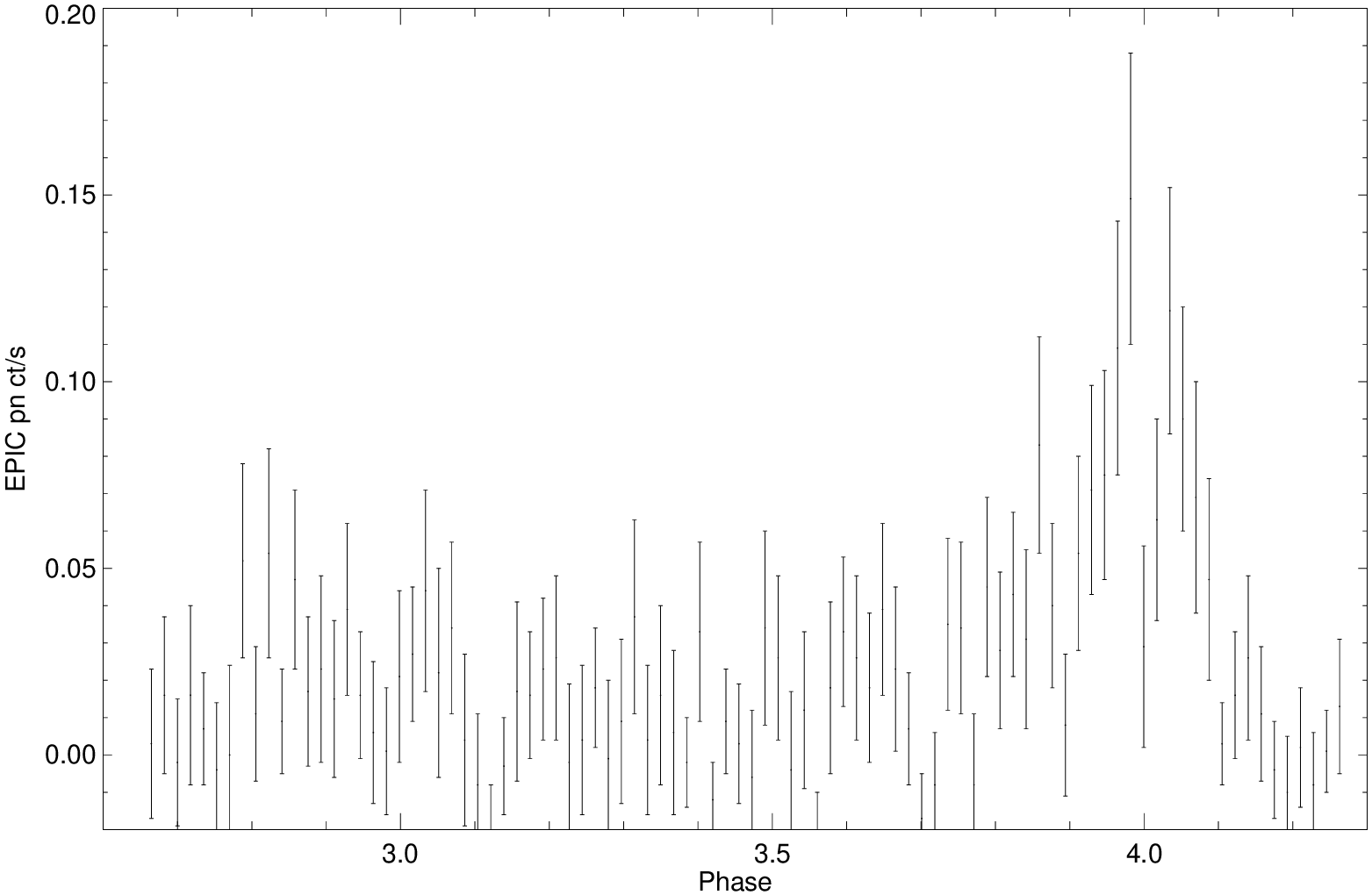}}
\put(4.5,-0.5){\includegraphics{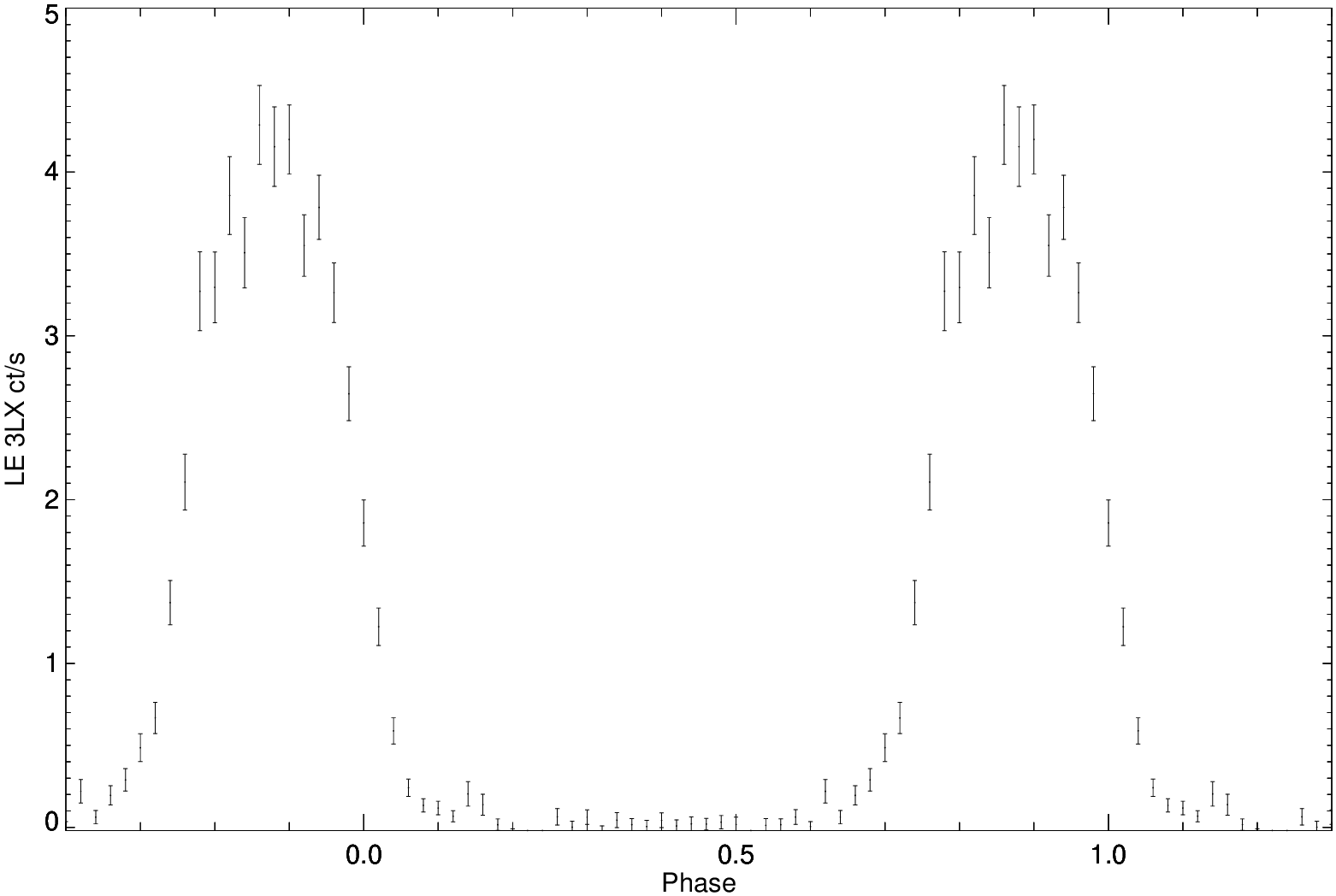}}
\end{picture}
\end{center}
\caption{X-ray observations of ST LMi. Left panel: {\sl XMM-Newton}
observations, Right panel: {\sl EXOSAT} observations.}
\label{rc_stlmi} 
\end{figure}

Although in low accretion states, a number of system showed evidence
for a modulation in the X-ray light curves indicating that accretion
was occuring at a low level. One system, ST LMi, showed evidence for
accretion occuring during one orbital cycle, but not the previous
cycle. The phasing is consistent with accretion occuring onto the main
accretion pole that has been seen to be X-ray bright at previous
epochs (Figure 5). This spectrum could be modelled with a
two-temperature thermal plasma model and the luminosity was
$\sim1\times10^{30}$ erg s$^{-1}$. This is similar with results of
{\sl XMM-Newton} observations of UZ For which also showed an accretion
event (Still \& Mukai 2001, Pandel \& Cordova 2002).

\vspace{1cm}
\noindent \underline{References}\\
Beuermann, K., Burwitz, V., 1995, ASP Conf Ser, 85, 99\newline
Cropper, M., et al, 1999, MNRAS, 306, 684\newline
Heise, J., Verbunt, F., 1988, A\&A, 189, 112\newline
Kuijpers, J., Pringle, J. E., 1982, A\&A, 114, L4\newline
Lamb, D. Q.,, Masters, A. R., 1979, ApJ, 234. L117\newline 
Pandel, D., et al, 2002, MNRAS, 332, 116\newline
Pandel, D., Cordova, F., 2002, MNRAS, 336, 1049 \newline
Ramsay, G. et al, 1994, MNRAS, 270, 692\newline
Ramsay, G., \& Cropper, M., 2002, MNRAS, 334, 805\newline
Ramsay, G., \& Cropper, M., 2003, MNRAS, 338, 219\newline
Schwope, A., et al, 2002, A\&A, 392, 541\newline
Still, M., \& Mukai, K., 2001, ApJ, 562, L71\newline

\end{document}